\documentclass{./iau} 
\usepackage{graphicx}

\title[African VLBI Network] 
{High-resolution radio astronomy: \\
an outlook for Africa}

\author[L.I. Gurvits et al.]   
{Leonid I. Gurvits$^{1,2,3}$,
 Robert Beswick$^4$, 
 Melvin Hoare$^5$, 
 Ann Njeri$^6$,
 Jay Blanchard$^{7,1}$,
 Carla Sharpe$^8$, 
 Adrian Tiplady$^8$,
 Aletha de Witt$^8$
  }

\affiliation{$^1$Joint Institute for VLBI ERIC, \\ Oude Hoogeveensedijk 4,
7991~PD Dwingeloo, The Netherlands \\ email: {\tt lgurvits@jive.eu} \\[\affilskip]
$^2$Dept. of Astrodynaics \& Space Missions, Delft University of Technology, \\ Kluyverweg 1, 2629 HS Delft, The Netherlands \\[\affilskip]
$^3$CSIRO Astronomy and Space Science, PO Box 76, Epping, NSW 1710, Australia \\[\affilskip]
$^4$ Jodrell Bank Centre for Astrophysics, The Alan Turing Building, Department of Physics and Astronomy, Oxford Road, The University of Manchester, M13 9PL, UK \\[\affilskip]
$^5$ School of Physics \& Astronomy, University of Leeds, Leeds, LS2 9JT, UK \\[\affilskip]
$^6$ Department of Physics and Astronomy, Oxford Road, The University of Manchester, M13 9PL, UK \\[\affilskip]
$^7$ National Radio Astronomy Observatory, PO Box O, 1003 Lopezville Rd., Socorro, NM 87801, USA \\[\affilskip]
$^8$ South African Radio Astronomy Observatory, 2 Fir Street, Black River Park, Cape Town, 7925, South Africa
}
\pubyear{2020}
\volume{356}  
\jname{Nuclear Activity in Galaxies Across Cosmic Time}
\editors{M. Povic, J. Masegosa, H. Netzer, P. Marziani, \\ 
P. Shastri, S.B. Tessema \& S.H. Negu, eds.}
\begin{document}

\maketitle

\begin{abstract}
Very Long Baseline Interferometry (VLBI) offers unrivalled resolution in studies of celestial radio sources. The subjects of interest of the current IAU Symposium, the Active Galactic Nuclei (AGN) of all types, constitute the major observing sample of modern VLBI networks. At present, the largest in the world in terms of the number of telescopes and geographical coverage is the European VLBI Network (EVN), which operates under the``open sky'' policy via peer-reviewed observing proposals. Recent EVN observations cover a broad range of science themes from high-sensitivity monitoring of structural changes in inner AGN areas to observations of tidal eruptions in AGN cores and investigation of redshift-dependent properties of parsec-scale radio structures of AGN. All the topics above should be considered as potentially rewarding scientific activities of the prospective African VLBI Network (AVN), a natural ``scientific ally” of EVN. This contribution briefly describes the status and near-term strategy for the AVN development as a southern extension of the EVN-AVN alliance and as an eventual bridge to the Square Kilometre Array (SKA) with its mid–frequency core in South Africa.  

\end{abstract}

\firstsection 
\section{Introduction}

High angular resolution studies of Active Galactic Nuclei (AGN) constitute a sizeable fraction of the science topics of the current IAU Symposium No.~356. In radio domain, these studies are conducted by using Very Long Baseline Interferometry (VLBI) systems spread over the entire surface of Earth and extended to Space. The `imaging angular resolution of VLBI reaches record high values of tens of microarcseconds as demonstrated recently in the ground-based observations of the Event Horizon Telescope (EHT) at 1.5~mm (\cite{EHT1-2019}) and in the Space VLBI observations by the RadioAstron mission at 1.3~cm (\cite{Bruni+2020}, and references therein). The geometry of a VLBI system is one of its key characteristics: in addition to the size (which does matter in interferometry), a distribution of interferometric elements defines the quality of images obtained by such a system. 

For almost the entire history of VLBI studies which began in 1967 (\cite{Moran1998}), the African continent was present in the VLBI world only by the Hartebeesthoek Observatory in South Africa (\cite{Gaylard+Nicolson-2007}). This single VLBI station on the entire continent provided an important baseline extension for the European and global VLBI Networks (EVN)\footnote{https://www.evlbi.org/home, accessed 2019.11.17}, the Australian Long Baseline Array (LBA)\footnote{https://www.atnf.csiro.au/vlbi/overview/index.html, accessed 2019.11.17} and other networks requiring a VLBI station in the Southern hemisphere. However, the large ``telescope-free'' area between Hartebeesthoek and Eurasia was always seen as a potentially attractive region for placing VLBI in the interests of various science tasks. 

A new momentum for VLBI in Africa came on the wave of the Square Kilometre Array (SKA) developments. The SKA, with its mid-frequency core in South Africa, has VLBI as one of its science-driven operational modes (\cite{Paragi+2015}). A natural synergy between developing the SKA as such and its partner observatories throughout the continent was realised at the first African SKA Partnership meeting at the Hartebeesthoek Observatory in 2003. Over the following years, three major components of the pan-African cooperation were formulated: development of the African VLBI Network (AVN), human capital development in areas related to radio astronomy, and the SKA governance. These components got a strong political support in the Pretoria Resolutions of 2014 and Memorandum of Understanding and Joint Development Plan of 2017.

\begin{figure}[t]
\begin{center}
 \includegraphics[width=9.0cm]{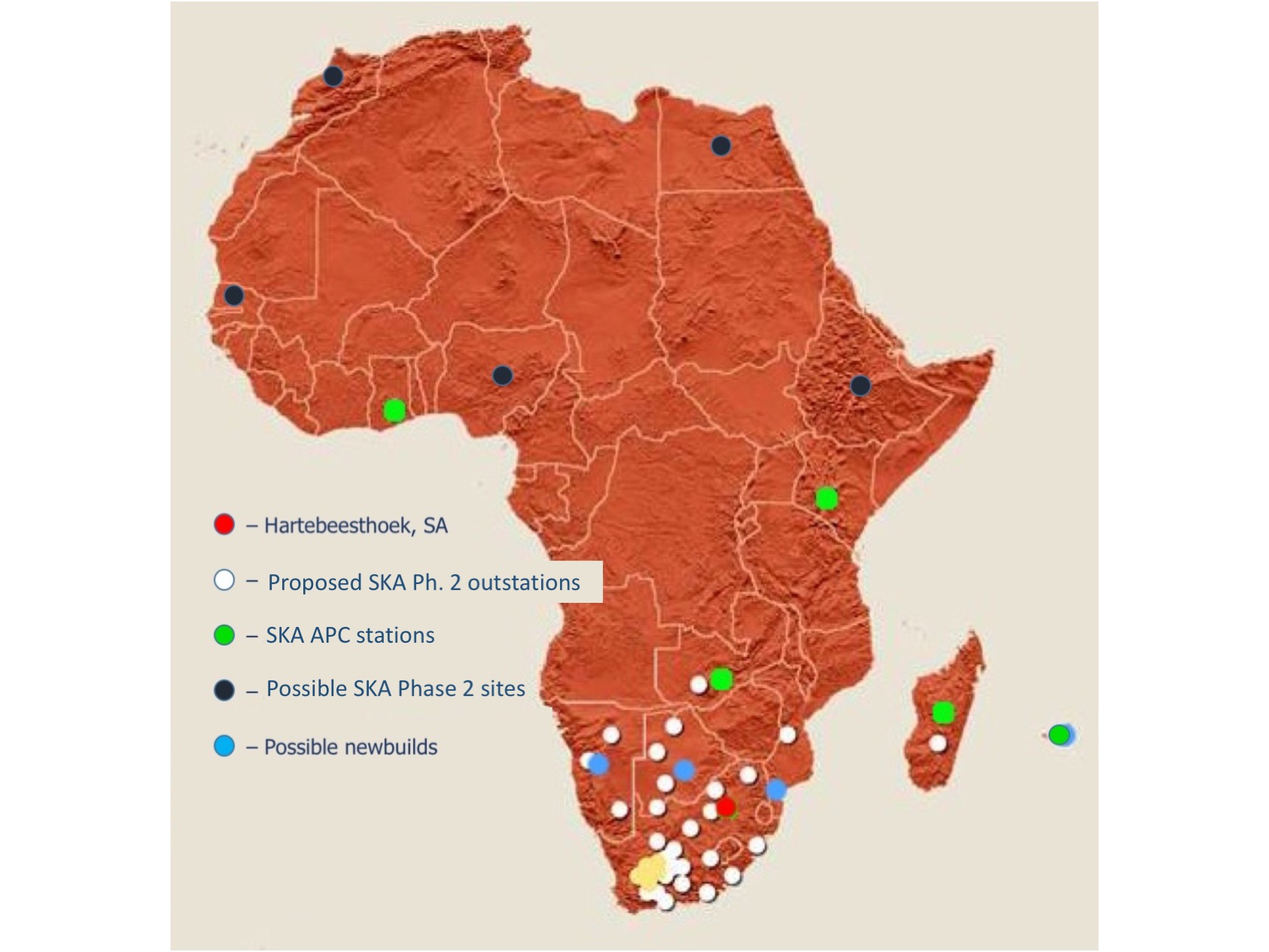} 
 \caption{Geography of prospective African VLBI Network radio telescopes. The red spot indicates the existing Hartebeesthoek VLBI telescopes, yellow -- the SKA core, white --  SKA outstations. Green dots indicate prospective AVN stations in SKA Africa Partner Countries (APC), and black and blue dots -- AVN stations under further considerations, possibly as SKA Phase 2 sites.}
   \label{fig-AVNmap}
\end{center}
\end{figure}

\section{The African VLBI Network}
     
A typical 30–m class VLBI radio telescope has a great deal of common technical features with a generic space communication antenna. Not surprisingly, many modern VLBI telescopes were designed or even built originally as Earth-Space communication facilities. The already mentioned first VLBI station in Africa, the 26~m radio telescope at the Hartebeesthoek Observatory in South Africa has began its life in 1961 as a NASA tracking station (\cite{Gaylard+Nicolson-2007}). More recently, a former Soviet military 32~m antenna in Irbene, Latvia, has become a radio telescope involved, among other science applications, in VLBI observations as a member of the EVN (\cite{Upnere+2013}). A 32~m satellite communication antenna at Warkworth in New Zealand has also become a radio telescope which, in particular, conducts LBA and global VLBI observations (\cite{Woodburn+2015}). The latter example is of particular interest for the African VLBI Network as the most relevant example.

\begin{figure}[h]
\begin{center}
 \includegraphics[width=13.4cm]{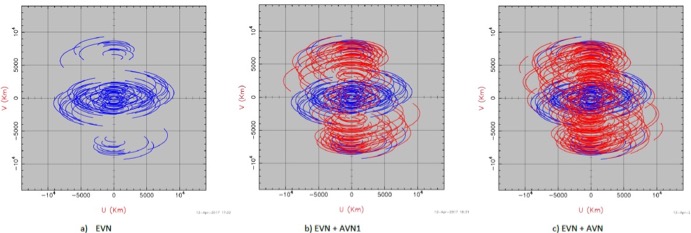}
 \includegraphics[width=13.4cm]{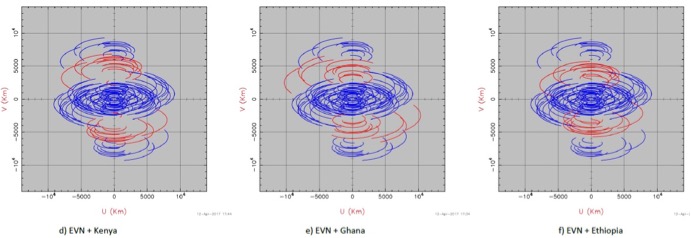}
 \caption{An example of a ``full–track'' (i.e., from rise to set) $uv$-coverage of a source at the declination $+20^{\circ}$. Blue tracks show EVN-only baselines, red tracks -- baselines to AVN antennas. The upper raw, from left to right, shows the $uv$-coverage for EVN-only, EVN plus the first wave of AVN, and EVN plus a full AVN. The bottom raw illustrates addition to the EVN single antennas in Kenya, Ghana and Ethiopia, respectively.}
   \label{AVN-uv}
\end{center}
\end{figure}

In the first decade of the 21st century, the global communication infrastructure has shifted its main data transport to fibre-optical cables. This shift released from active duty many satellite communication antennas. A census of such facilities defined about a dozen of decommissioned or nearly-decommissioned 30-m-class antennas distributed over the African continent. Fig.~\ref{fig-AVNmap} presents the expected geographical distribution of the prospective AVN telescopes. The deployment of AVN is seen as a two-stage process, with the first wave of telescopes, led by the radio astronomy cluster in South Africa and 32-m antennas in Ghana, Kenya, Madagascar, Mauritius, Namibia and Zambia. These telescopes will be followed by antennas in Egypt, Ethiopia. Morocco, Nigeria and Senegal.

The geometry of an interferometer is one of its main characteristics. Ideally, the geometry should be optimised using complex criteria describing the quality of reconstructed images under various limitations and boundary conditions. Such the optimisation was exercised, e.g., for the dedicated VLBI network -- the Very Long Baseline Array (VLBA, \cite{Napier+1994}) and the SKA (\cite{Lal+2010}, and references therein). Due to the ad hoc approach to the formation of the AVN, such the optimisation is obviously impossible -- antennas' locations should be taken as a given. Nevertheless, an analysis of the AVN configuration is a necessary component of the overall project as it allows us to evaluate the observational capability of the network. For an interferometric array, the most straightforward representation of its observational potential is given by the 2D distribution of sampling of the spectra of spatial frequencies, the so-called $uv$-coverage. Fig.~\ref{AVN-uv} illustrates the value of new AVN stations in combination with the existing EVN stations. Even single additional antennas (illustrated in the bottom raw) provide a sensible ``filling-in'' improvement in the $uv$-gap at the north-south intermediate baselines. The configuration consisting from full EVN and AVN arrays offers a nearly perfect $uv$-coverage.   

\begin{figure}[!tbp]
  \centering
  \begin{minipage}[b]{0.46\textwidth}
    \includegraphics[width=\textwidth]{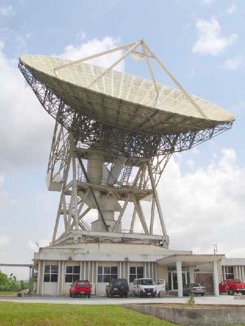}
   \end{minipage}
  \hfill
  \begin{minipage}[b]{0.42\textwidth}
    \includegraphics[width=\textwidth]{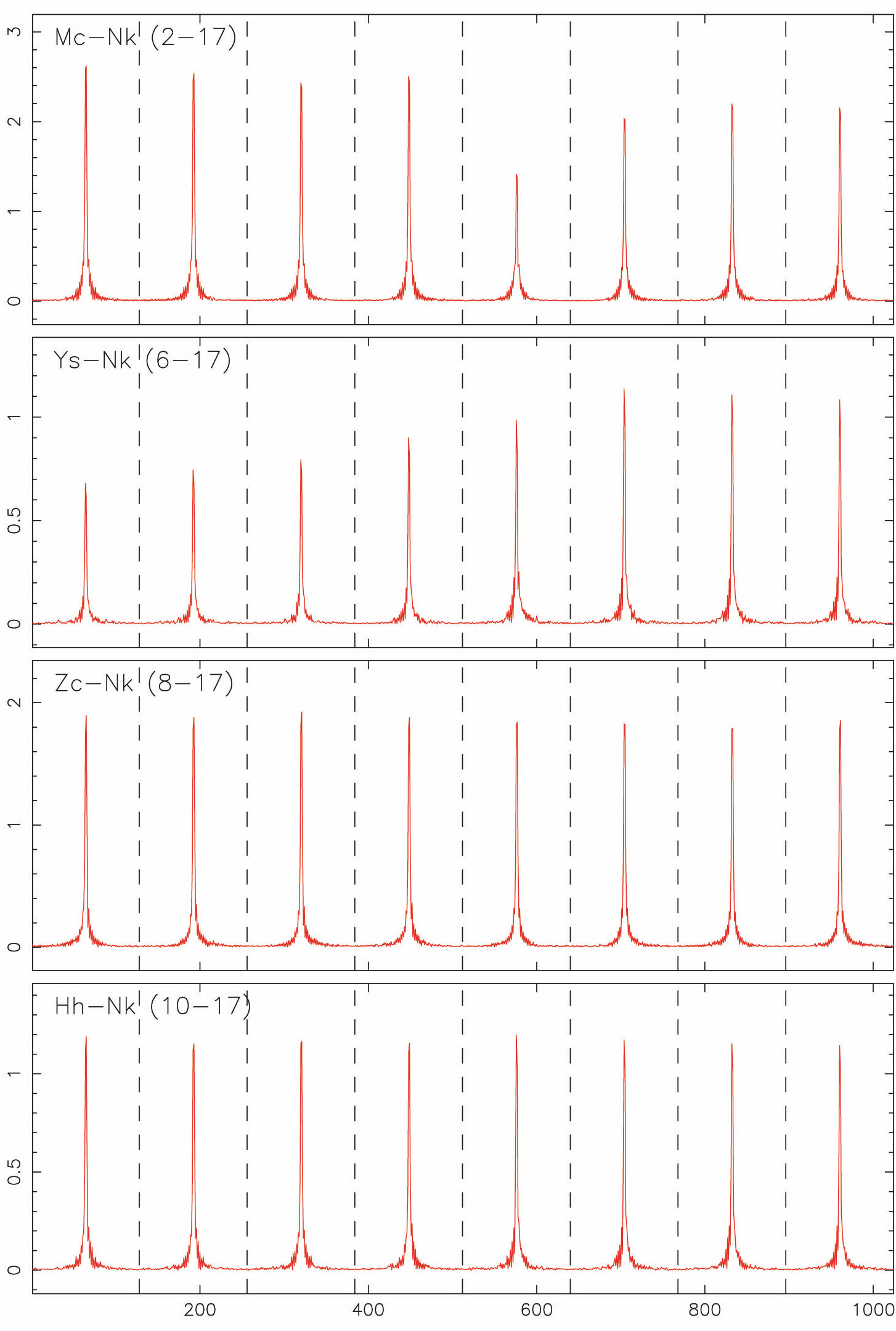}
  \end{minipage}
  
  \caption{{\it Left}: The Kuntunse 32~m satellite communication antenna, Ghana. {\it Right}:  The first VLBI fringes to the Kuntunse telescope on baselines to the telescopes (top--down) Medicina (Italy), Yebes (Spain), Zelenchukskaya (Russia) and Hartebeesthoek (South Africa).}
\label{fig-Kuntunse}   
\end{figure}

\section{Kuntunse radio telescope, Ghana}

The Ghana Radio Astronomy Observatory in Kuntunse (also spelled sometimes as Kutunse or Nkuntunse), located about 25~km north-west of the capital city of Accra, is the first AVN observatory undergoing extensive refurbishment of its instrumentation (\cite{Asabere+2015}). Its 32–m antenna (Fig.~\ref{fig-Kuntunse}, left panel) operated as a telecommunication facility between 1981 and 2008. In 2011, the antenna's ownership was transferred to the newly established Ghana Space Science and Technology Institute (GSSTI). In collaboration with the SKA--SA, now known as the South African Radio Astronomy Observatory (SARAO) and EVN institutes, the GSSTI staff prepared the facility for the first test VLBI observations in  2017. The test observation was conducted with the C-band (3.8 -- 6.4~GHz) communication receiver. Fig.~\ref{fig-Kuntunse} (right panel) presents the clear VLBI fringes obtained at JIVE on baselines between Kuntunse and several telescopes in Europe and South Africa. 

\section{Educational potential of AVN}

\begin{figure}[h]
\begin{center}
 \includegraphics[width=8.8cm]{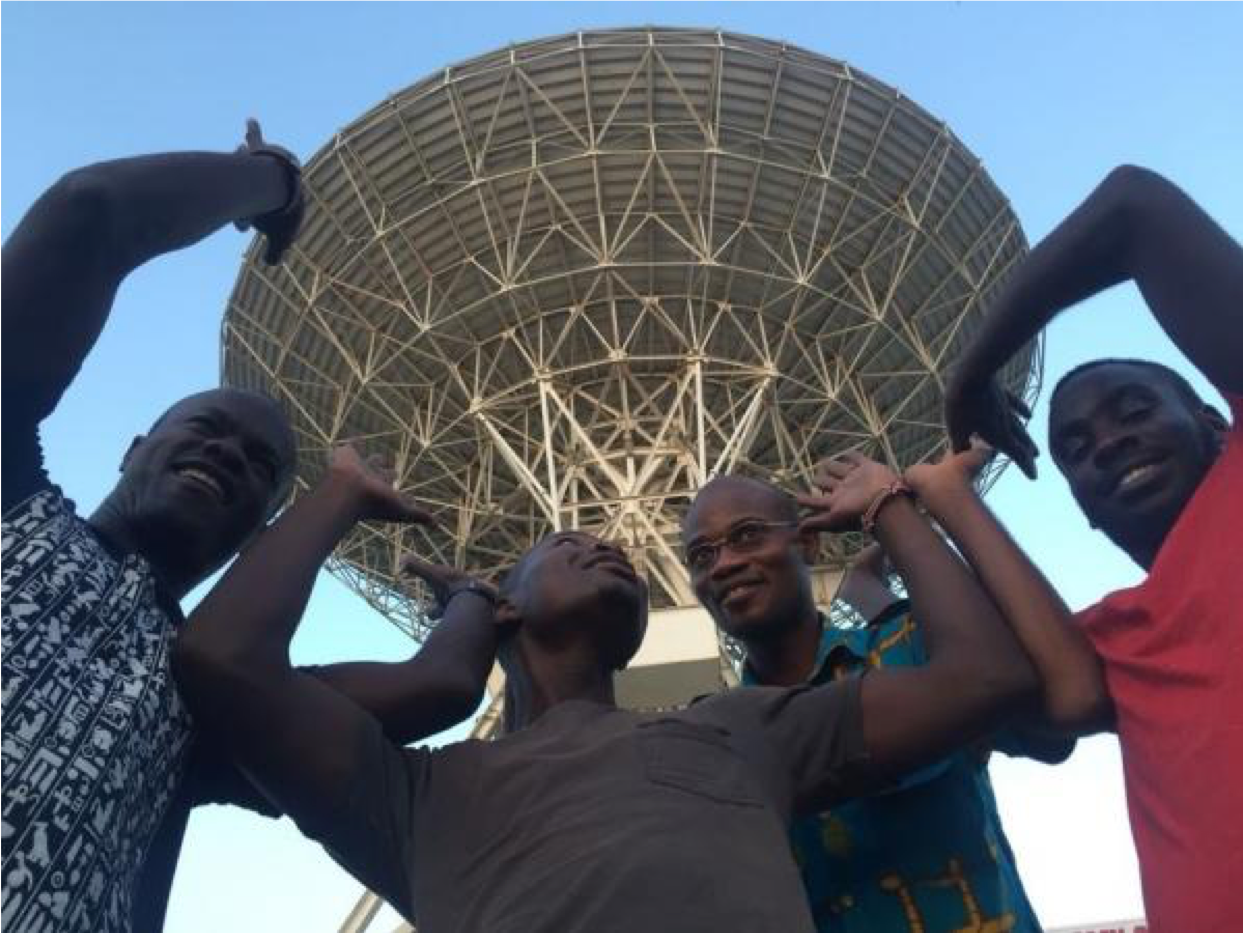} 
 \caption{Participants of the radio astronomy school at Kuntunse, organised jointly by the JUMPIG JIVE and DARA projects, May 2018.}
   \label{fig-Kuntunse-school}
\end{center}
\end{figure}

The radio astronomy infrastructure in Africa and in particular AVN-related activities serve as an efficient setting for educational and public outreach activities on the continent. A number of joint schools (e.g., a school in Kuntunse, May 2018, Fig,~\ref{fig-Kuntunse-school}) workshops and exchange visits conducted in the past several years assist in building up the professional radio astronomy community around the prospective AVN observatories. Hopefully, at the successive symposium in Addis Ababa of 2024, young researchers from African institutes will presets studies conducted with the African VLBI Network.

{\bf Acknowledgements}

Collaborative activities with the prospective African VLBI Network are supported by the European Commission Horizon 2020 Research and Innovation Programme under grant agreement No.~ 730884 (JUMPING JIVE). The DARA project is funded by the UK's Newton Fund via grant ST/R001103/1 from the Science and Technology Facilities Council and by South Africa's National Research Foundation. LIG expresses gratitude to the Leids Kerkhoven-Bosscha Fonds for partial support of participation in the IAU Symposium No.~356.


\begin{thebibliography}{}

\bibitem[Asabere et al. 2015]{Asabere+2015}
{Asabere, B.D., Gaylard, M.J., Horellou, C., et al.} 2015,
arXiv:1503.08850

\bibitem[Bruni et al. 2019]{Bruni+2020} 
{Bruni, G., Savolainen, T., G\'{o}mez, J.L., et al.} 2020,
\textit{Adv. Sp. Res.}, 65(2), 712

\bibitem[EHT 2019]{EHT1-2019}
{Event Horizon Telescope Collaboration} 2019,
\textit{ApJL}, 875, L1 

\bibitem[Gaylard \& Nicolson 2007]{Gaylard+Nicolson-2007} 
{Gaylard, M.J. \& Nicolson, G.D.} 2007,
\textit{African Sky}, 11, 49

\bibitem[Lal et al. 2010]{Lal+2010}
{Lal, D.V., Lobanov, A.P., \& Jim\'{e}nez-Monferrer, S.} 2010, 
arXiv:1001.1477

\bibitem[Moran 1998]{Moran1998}
{Moran, J.M.} 1998, 
in J.A. Zensus, G.B. Taylor \& J.M. Wrobel (eds.) 
\textit{Radio Emission from Galactic and Extragalactic Compact Sources}, IAU Coll. 164, ASP Conf. Series, 144, 1

\bibitem[Napier et al. 1994]{Napier+1994} 
{Napier, P.J., Bagri, D.S., Clark, B.G., et al.} 1994, 
\textit{Proc. IEEE}, 82, No.~5, 658 

\bibitem[Paragi et al. 2015]{Paragi+2015}
{Paragi., Z., Godfrey, L., Reynolds, C. et al.} 2015,
in \textit{Advancing Astrophysics with the Square Kilometre Array}, PoS(AASKA14)143

\bibitem[Upnere et al. 2013]{Upnere+2013}
{Upnere, S., Jekabsons, N., \& Joffe, R.} 2013, 
\textit{J. of Theoretical and Applied Mechanics}, 43, 39

\bibitem[Woodburn et al. 2015]{Woodburn+2015}
{Woodburn, L., Natusch, T., Weston, et al.} 2015, 
\textit{PASA}, 32, e017

\end{thebibliography}
\end{document}